\documentclass[aps,prl,twocolumn,showpacs,superscriptaddress]{revtex4}

\usepackage{graphicx}
\usepackage{dcolumn}
\usepackage{bm}
\usepackage{amssymb}
\usepackage{color}

\begin{document}


\title{Confinement-induced shape transitions in multilamellar vesicles}

\author{Ai Sakashita}
\affiliation{Department of Physics, Ochanomizu University, 2-1-1 Otsuka, Bunkyo, Tokyo 112-8610, Japan}
\affiliation{
Institute for Solid State Physics, University of Tokyo,
 Kashiwa, Chiba 277-8581, Japan}
\author{Masayuki Imai}
\affiliation{Department of Physics, Faculty of Science, Tohoku University, 6-3 Aramaki, Aoba-ku, Sendai, Miyagi 980-8578, Japan}
\author{Hiroshi Noguchi}
\email[]{noguchi@issp.u-tokyo.ac.jp}
\affiliation{
Institute for Solid State Physics, University of Tokyo,
 Kashiwa, Chiba 277-8581, Japan}

\date{\today}

\begin{abstract}
Morphologies of a vesicle confined in a spherical vesicle were explored experimentally by fast confocal laser microscopy
and numerically by a dynamically-triangulated membrane model with area-difference elasticity.
The confinement was found to induce several novel shapes of the inner vesicles, that had been never observed in unilamellar vesicles:
double and quadruple stomatocytes, slit vesicle, and vesicles of two or three compartments with various shapes.
The simulations reproduced the experimental results very well and
some of the shape transitions can be understood by a simple theoretical model for axisymmetric shapes.
\end{abstract}

\maketitle

Cells and cell organelles have various shapes depending on their functions.
For example, red blood cells (RBC) have a biconcave disk shape; this
allows large deformations with a fixed area and volume so that RBCs can flow in
microvessels narrower than themselves \cite{fung04,fedo13}. 
These discocyte shapes can be observed in  unilamellar liposomes and
reproduced by  minimizing the membrane bending energy with area and volume constraints \cite{canh70,helf73}.
In living cells,  organelles such as Golgi apparatus, endoplasmic reticulum, and mitochondria
have much more complicated shapes;
their local curvatures are considered to be regulated by BAR and other proteins \cite{shib09,baum11}.
Among these  organelles,
mitochondria have a specific feature, {\it i.e.}, it consists of two bilayer membranes \cite{frey00,mann06,sche08}.
The inner membrane has a much larger surface area than the outer one and forms numerous invaginations called cristae.
The cristae have tubular and planar structures, and their narrow junctions regulate diffusion between different compartments.
Although the confinement by the outer membrane is expected to play a role in determining the shape of the inner membrane,
it is not well understood so far.
In this letter, we reveal the effects of the confinement using multilamellar liposomes as a simple model system.

Unilamellar liposomes form various morphologies such as stomatocyte, pear, pearl-necklace, and branched starfish-like 
shapes.
All of these shapes can be reproduced by the area-difference-elasticity (ADE) model \cite{lipo95,svet89,seif97,khal08,saka12}.
In contrast, the shapes of  multilamellar vesicles have not received attention.
The main reasons are experimental difficulties with distinguishing inner and outer membranes and controlling the volume and area of inner vesicles.
We used a fast confocal laser microscope to extract three-dimensional (3D) images of multilamellar liposomes and observed several confinement-induced shapes of the inner vesicles.

Kahraman {\it et al.} \cite{kahr12a, kahr12b} recently simulated a vesicle in spherical or ellipsoidal confinement
and found that weak confinement produces a stomatocyte with an open (circular or elliptic) neck;
 further confinement results in the formation of a double stomatocyte where the inner sphere of a typical stomatotye is filled 
by additional spherical invagination.
Hereafter, we call a normal stomatocyte a single stomatocyte
to distinguish it from a double stomatocyte.
Kahraman {\it et al.} considered the bending energy with weak spontaneous curvatures but not the ADE energy.
We simulated the confined vesicles by using a dynamically-triangulated membrane with the ADE model
and obtained many more varieties of vesicle shapes, including experimentally observed shapes.
We also analyzed the mechanism for shape determination using a simple theoretical model of axisymmetric shapes.

We prepared single-component vesicles from DOPC (1,2-dioleoyl-sn-glycero-3-phosphocholine, Avanti Polar Lipids) 
using the gentle hydration method with pure water \cite{saka12}. 
TR-DHPE (Texas Red, 1,2-dihexadecanoyl-sn-glycero-3-phosphoethanolamine, Molecular Probes) was used as the dye. 
We kept vesicle suspensions at room temperature ($24$-$25^\circ$C) and observed them by a fast confocal laser microscope (Carl Zeiss, LSM 5Live). 
At this stage, most vesicles formed either a spherical or tubular shape spontaneously, and some had one or more vesicles inside. 
We then focused on multilamellar vesicles.

Various shapes of the inner liposomes were observed (see Figs. \ref{fig:sto2db}(a) and \ref{fig:phase_v}(a)  and the 3D image of a double stomatocyte in Supplemental Material \cite{epaps1}).
Among them, the
double stomatocytes [Fig. \ref{fig:phase_v}(a)iv] and open single stomatocytes [Fig. \ref{fig:sto2db}(a)i]
agree with the shapes predicted in the previous simulations {~\cite{kahr12a, kahr12b}}.
We also observed other liposome shapes that have not been previously reported: 
a vesicle with a deep slit (Fig. \ref{fig:sto2db}(a)iii,iv),
two hemispheres (or buds) connected by a small neck [called doublet, 
see Figs. \ref{fig:sto2db}(a)v and \ref{fig:phase_v}(a)ii],
and a single stomatocyte of discoidal invagiation [see Fig. \ref{fig:phase_v}(a)iii].

\begin{figure}
\includegraphics{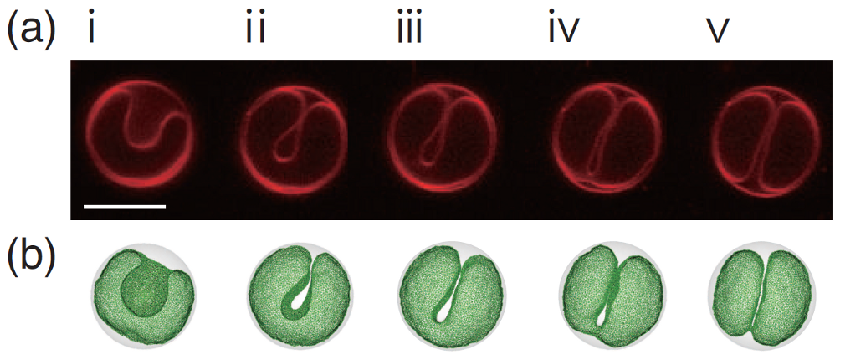}
\includegraphics{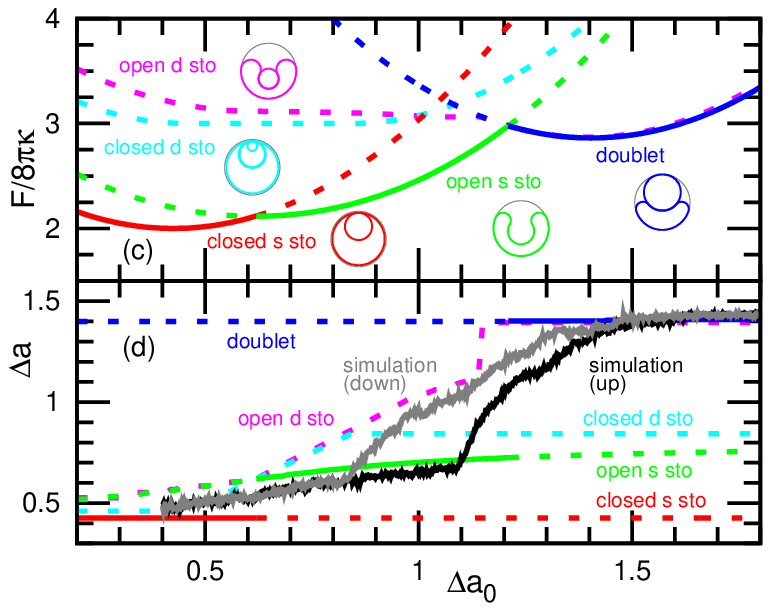}
\caption{
Shape changes of liposome confined in spherical liposome.
(a) Time-sequential microscopy images of liposomes.
From left to right: $t=0$, $223$, $225$, $227$, and $230$ s.
Scale bar: $10$ $\mu$m.
(b) Sequential snapshots of triangulated-membrane simulation
at  $v_{\rm r}=0.6$ and $v_{\rm {con}}= 0.727$.
Intrinsic area difference $\Delta a_0$ gradually increased.
From left to right: $\Delta a_0=0.93$, $1.30$, $1.44$, $1.55$, and $1.80$.
Inner and outer vesicles are shown in green and light gray, respectively.
(c) Free energy $F$ of simple axisymmetric shapes.
Solid and dashed lines represent energy-minimum and metastable states, respectively.
Typical shapes are shown in the inset.
(d) Area difference $\Delta a$ of two leaflets depending on $\Delta a_0$.
Black and gray lines represent simulation results with increasing or decreasing  $\Delta a_0$,
respectively.
Other solid and dashed lines represent same data in (c) for simple axisymmetric shapes.
}
\label{fig:sto2db}
\end{figure}

In order to understand these distinctive vesicle shapes,
we describe the inner vesicle as a fluid membrane of volume $V_{\rm {ves}}$ and outer vesicle as a hard sphere of volume $V_{\rm {sp}}=(4\pi/3) R_{\rm {sp}}^3$.
In the ADE model, the free energy of a single-component vesicle with a fixed topology is given by
\begin{equation}
F =  \int  \Bigl[ \frac{\kappa}{2}(C_1+C_2)^2  \Bigr]  dA
                       + \frac{\pi k_{\rm {ade}}}{2Ah^2}(\Delta A - \Delta A_0)^2
\label{eq:ade}
\end{equation}
where $C_1$ and $C_2$ are the principal curvatures at each point 
in the membrane.
The coefficients $\kappa$ and $k_{\rm {ade}}$
are the bending rigidity and ADE coefficient, respectively.
The areas of the outer and inner leaflets of a bilayer vesicle
differ with  $\Delta A= h \oint (C_1+C_2) dA$,
where $h$ is the distance between the two leaflets.
Since the flip-flop of lipids (traverse motion between leaflets)
is very slow, the area difference $\Delta A_0=(N_{\rm {out}}-N_{\rm {in}})a_{\rm {lip}}$
preferred by lipids is typically different from  $\Delta A$,
where $N_{\rm {out}}$ and $N_{\rm {in}}$
are the number of lipids in the outer and inner leaflets, respectively,
and $a_{\rm {lip}}$ is the area per lipid.
We used typical values for the lipid membranes $\kappa=k_{\rm {ade}}=20k_{\rm B}T$,
where $k_{\rm B}T$ is the thermal energy \cite{seif97,saka12}.
We present our results using the reduced volume $v_{\rm r}= V_{\rm {ves}}/(4\pi R_{\rm A}^3/3)$,
 confinement volume ratio $v_{\rm {con}}= V_{\rm {ves}}/V_{\rm {sp}}$,
  reduced area differences $\Delta a =\Delta A/8\pi h R_{\rm A}$, and $\Delta a_0 = \Delta A_0/8\pi h R_{\rm A}$,
where $R_{\rm A}= \sqrt{A/4\pi}$.

\begin{figure}
\includegraphics{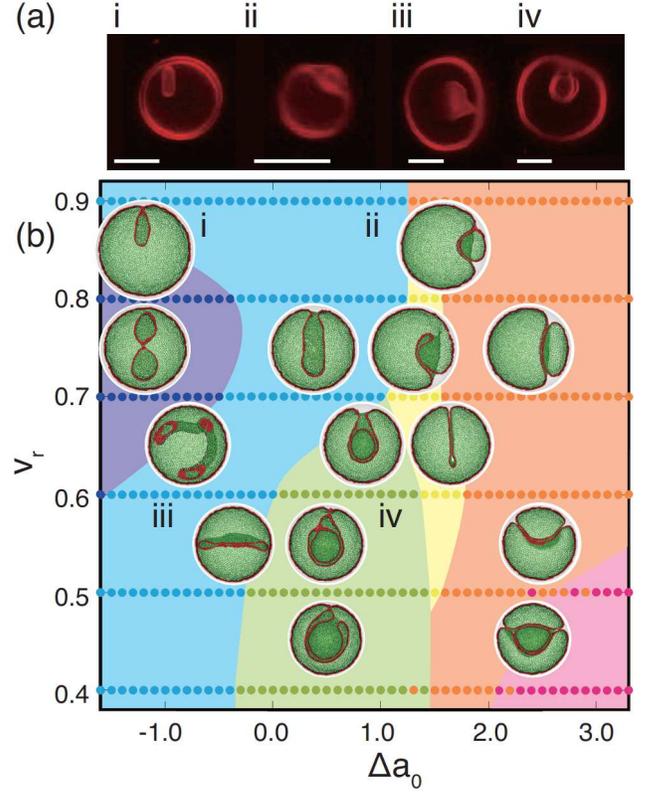}
\caption{
(a) Microscopy images of liposomes with various shapes of inner liposomes.
i) Single stomatocyte of prolate invagiation. ii) Doublet. iii) Single stomatocyte of discoidal invagiation. 
iv) Double stomatocyte.
Scale bar: $10$ $\mu$m.
(b) Phase diagram at the most confined state in simulations on ($v_{\rm r}, \Delta a_{0}$) plane. 
Shapes are categorized by types of compartmentalization:
(blue) single stomatocytes of one inner bud, (violet) single stomatocytes of two inner buds,
(green) double stomatocytes, (yellow) slit vesicles, (orange) doublets, 
and (pink) triplets.
Typical simulation snapshots are shown in the inset of (b).
Their front halves are removed and cut-off sections are  shown in red.
}
\label{fig:phase_v}
\end{figure}

The double stomatocyte can be modeled as three spheres (radii $R_1 \ge R_2\ge R_3$) connected by two small necks.
The volume, area, and free energy $F$ are estimated as $V_{\rm {ves}}=(4\pi/3)(R_1^3- R_2^3+ R_3^3)$, $A=4\pi(R_1^2+ R_2^2+ R_3^2)$,
and  $F/8\pi\kappa=3 + \pi\{ (R_1- R_2+ R_3)/R_{\rm A} - \Delta a_0\}^2$, respectively.
At $R_{\rm {sp}}=R_1$ and $R_2=R_3$, the internal space of the outer sphere is completely filled by the inner vesicle, {\it i.e.},
$v_{\rm {con}}=1$, and $F/8\pi\kappa= 3 + \pi( {v_{\rm r}}^{1/3} - \Delta a_0)^2$ for $v_{\rm r} \geq 3^{-3/2} \simeq 0.19$.

Dynamically-triangulated membrane models
are widely used to study the shape deformation of membranes and vesicles \cite{gomp04c,nogu09}.
We used the molecular dynamics of this membrane model with a Langevin thermostat \cite{nogu09}.
The membrane is described by vertices of $N_{\rm v}=4000$, and
the mean distance of the bonds connecting neighboring vertices is $\sigma= 0.06R_{\rm A}$.
Free energy $F$ in Eq.~(\ref{eq:ade}) is discretized 
using dual lattices of triangulation and
the area and volume are maintained by harmonic constraint potentials \cite{nogu05}.
The vertices have short-range excluded volume interactions with each other and 
with the outer sphere.
To obtain the most confined state,
the radius of outer sphere is varied as $R_{\rm {sp}}=R_{\rm {max}}+0.05\sigma$;
 the repulsion with the outer sphere  efficiently pushes the inner membranes into compact states,
where $R_{\rm {max}}$ is the maximum distance of the vertices from the center of the sphere.
Since the excluded volumes yield the finite minimum distance $l_{\rm {mb}}\sim \sigma$ between the membranes,
$v_{\rm {con}}$ does not reach unity.
The minimum radius $R_{\rm {sp}}^{\rm {min}}$ obtained in the simulation
agrees with the estimation by the three-sphere model using the finite minimum distance \cite{epaps1}.
The double stomatocyte is only obtained at $v_{\rm r} \lesssim 0.6$.
The details on the model are described in Supplemental Material \cite{epaps1}.

The simulations of the ADE model reproduced
all of the experimentally observed shapes [see Figs. \ref{fig:sto2db}(b) and \ref{fig:phase_v}(b)].
In particular, sequential shape transitions of a liposome from an open single stomatocyte to a doublet via a slit vesicle were 
 reproduced by gradual changes to the intrinsic area difference  $\Delta a_0$ (compare Figs. \ref{fig:sto2db}(a) and (b) 
and see a movie in Supplemental Material \cite{epaps1}).
Previously, we observed similar $\Delta a_0$ changes in unilamellar liposomes under the same experimental conditions \cite{saka12}.
Small reservoirs of lipid are likely present on the membrane, and
the laser illumination of the microscopy induces  fusion into either leaflet, which leads to changes in $\Delta a_0$.
In the unilamellar (unconfined) liposomes, as $\Delta a_0$ increases, a single stomatocyte changes into 
a discocyte and then to starfish or pearl-necklace shapes.
The confinement destabilizes the discocyte shape; instead,
transient open single stomatocytes become stable.
The doublets obtained in the confinement are understood as compressed  pearl-necklace shapes.
The slit vesicle shape is not seen in unilamellar liposomes.
The enlargement of an ellipsoidal neck of the open stomatocyte
leads to the formation of a tongue-like invagination,
that resembles a crista shape in the old view of the classic baffle model \cite{frey00}.

\begin{figure}
\includegraphics{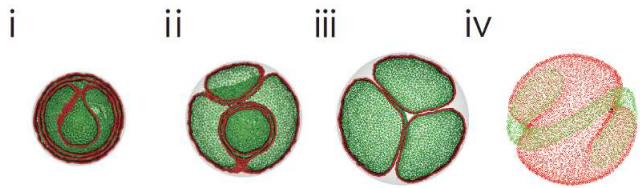}
\caption{(color online) Snapshots of vesicles in metastable states.
(i) Quadruple stomatocyte at $(v_{\rm r}, \Delta a_{0})=(0.15, 0.80)$. 
(ii) Double stomatocyte with outer bud at $(v_{\rm r}, \Delta a_{0})=(0.40, 1.40)$. 
(iii) Triplet at $(v_{\rm r}, \Delta a_{0})=(0.5, 2.3)$.
(iv)  Tubular (green) compartment wraps the other spherical (red) compartment at $(v_{\rm r}, \Delta a_{0})=(0.54, 3.8)$.
(i--iii) Half-cut snapshots are displayed.
}
\label{fig:others}
\end{figure}

\begin{figure}
\includegraphics{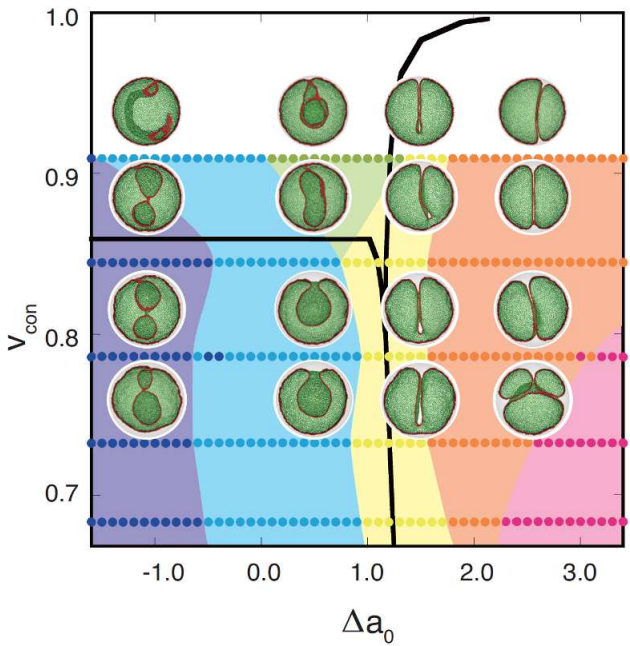}
\caption{Phase diagram of ($v_{\rm {con}}, \Delta a_{0}$) plane at $v_{\rm r}=0.6$.
Solid black lines represent phase boundaries estimated by the simple model for axisymmetric shapes.
Half-cut snapshots are displayed in the inset.
}
\label{fig:phase_r}
\end{figure}

Using the minimum radius $R_{\rm {sp}}^{\rm {min}}$ at  $\Delta a_0=0.4$,
the phase diagram is constructed as shown in Fig. \ref{fig:phase_v}(b).
Note that two or more shapes coexist (in particular, several shapes at $v_{\rm r}=0.4$),
so that the shape with the lowest energy is chosen.
Some of the metastable shapes are shown in Fig. \ref{fig:others}.
The obtained shapes are categorized by the number of compartments separated by necks.
As $\Delta a_0$ increases, the vesicle shapes change from inner-budded shapes (single stomatocyte) 
to outer-budded shapes (doublets and triplets).
In the middle range, slit vesicles and double stomatocytes appear.

Although we consider five categories for simplicity, as shown 
 in Fig. \ref{fig:phase_v}(b),
the shapes in each category have detailed variations.
At $v_{\rm r}=0.9$, 
the doublet consists of one small spherical compartment and one large compartment.
With decreasing $v_{\rm r}$,
the size of the smaller compartment increases and reaches the same size of the other at  $v_{\rm r}\simeq 0.6$.
At lower $v_{\rm r}$, two compartments become interdigitated like the seam of a baseball 
[see the second-lowest snapshot on the right of Fig. \ref{fig:phase_v}(b)].
As $v_{\rm r}$ decreases,
the inner bud of the single stomatocyte changes its shape:
 a sphere,  prolate, discocyte, and bent tube
 [see four snapshots in the blue region of Fig. \ref{fig:phase_v}(b)].
Since the radius of the single stomatocyte is restricted by the outer sphere,
the excess area to the confined volume does not always allow a spherical shape of the inner bud.
Thus, the inner bud exhibits similar shape transitions to  a unilamellar vesicle with 
the reduced volume $v_{\rm r}^{\rm {in}}=v_{\rm r}(1-v_{\rm {com}})/(v_{\rm {con}}^{2/3}-v_{\rm {r}}^{2/3})^{3/2}$
and area difference 
$\Delta a_{\rm {in}}= \{(v_{\rm r}/v_{\rm {con}})^{1/3}-\Delta a\}/\sqrt{1-(v_{\rm r}/v_{\rm {con}})^{2/3}}$. 
The discoidal invagination connected to spherical membranes via a narrow neck 
and tubular invagination
resemble the crista structures
 in the modern cristae model based on 3D electron microscopy observations \cite{frey00,mann06,sche08}.
The large surface area of the inner  mitochondrial membrane 
 and small volume between the inner and outer membranes
are key factors to determine the crista structures.

Around the right (left) boundary of the double-stomatocyte region in Fig. \ref{fig:phase_v}(b),
the neck connecting the outer (inner) and middle membranes opens.
As the vesicles cross the left phase boundary, 
the inner membranes exhibit a shape transition to a discocyte [snapshots (iv) to (iii) in Fig. \ref{fig:phase_v}(b)].
A typical triplet shape is a spherical compartment surrounded by two  compartments
[see bottom-right snapshot in Fig. \ref{fig:phase_v}(b)].
Three symmetric compartments (like sections of oranges) are obtained as a metastable state
[see Fig. \ref{fig:others}(iii)].
Quadruple stomatocytes can be formed at $v_{\rm r} \leq 0.15$ [see Fig. \ref{fig:others}(i)].

The effects of the confinement strength
were investigated through simulations that varied the outer radius $R_{\rm {sp}}$ (see Fig. \ref{fig:phase_r}).
With increasing confinement (increasing $v_{\rm {con}}$),
the number of compartments decreases at large values of $\Delta a_0$;
the triplets transform into doublets.
This is caused by larger deformation under greater confinement.
More deformed doublets can have larger values of  $\Delta a_0$.
The double stomatocyte appears when the confinement is sufficiently strong at $v_{\rm {con}} > 0.85$.

To understand the confinement effects more deeply, 
we analytically calculated the membrane free energy of several simple geometries:
single and double stomatocytes with open or closed necks and doublets.
Only axisymetirc shapes are considered,
and the cross-sectional shape is represented as a combination of arcs 
[see cross-section shapes drawn in the insets of Fig. \ref{fig:sto2db}(c)].
Details on the model and calculation are described in Supplemental Material \cite{epaps1}.
This model can well explain the shape transition from the single  stomatocyte to the doublet (see Fig. \ref{fig:sto2db}).
Although the slit vesicle is not considered because of its lack of axisymmetry,
$\Delta a$ dependence on $\Delta a_0$ is also well reproduced 
 [see Fig. \ref{fig:sto2db}(d)].
The doublet consisting of two hemispheres of the same size
is obtained not at $v_{\rm r}=0.6$ but at $v_{\rm r}=0.65$ in this simple model.
This difference from the experimental and simulation results
would be due to the limitation of this simple shape representation.

At the limit of the full confinement, {\it i.e.}, $v_{\rm {con}}\to 1$,
the double stomatocyte has the lowest energy at any $\Delta a_0$ (see the solid line in Fig. \ref{fig:phase_r}).
The other possible shapes have sharply-bent edges,
which bending energy diverges at $v_{\rm {con}}\to 1$.
Thus, the double stomatocytes are in thermodynamically stable state at $v_{\rm {con}}\to 1$
for $v_{\rm r} \geq 3^{-3/2}$.
For  $v_{\rm r} < 3^{-3/2}$, two inner buds of the double stomatocyte
 buckle together to form open or closed quadruple stomatocytes.

In summary,
we revealed various vesicle shapes induced by the confinement in a spherical vesicle.
Some of the shapes, such as double stomatocytes and slit vesicles, were observed experimentally for the first time.
All of the shapes were reproduced by the computer simulation when the ADE energy is taken into account.
Interestingly, tubular and discoidal invaginations obtained at low values of $\Delta a_0$
are very similar to the crista structures in mitochondria.
This may suggest that the ADE energy is one of the key quantities to determine the crista structures.
The outer vesicles were modeled as a sphere here, but
the outer membranes of liposomes and mitochondria are also flexible and can exhibit tubular or more complicated shapes.
The effects of deformation of the outer membrane and local spontaneous curvatures induced by proteins
are an important unsolved problem for future studies.

\begin{acknowledgments}
We thank P. Ziherl for informative discussions.
This work was partially supported by the Japan Society for the Promotion of Science
and the female leadership program at Ochanomizu University.
\end{acknowledgments}


\end{document}